 \documentstyle[12pt]{article}
 
\begin{document}

 \topmargin 0pt
 \oddsidemargin 5mm
 \begin{titlepage}
 \setcounter{page}{0}
 \rightline{Preprint YERPHY-1442(12)-95}

 \vspace{2cm}
 \begin{center}
 {\Large Pseudoclassical neutrino in the
 external electromagnetic field}
 \vspace{1cm}

 {\large Grigoryan G.V.,  Grigoryan R.P.} \\
 \vspace{1cm}
 {\em Yerevan Physics Institute,  Republic of Armenia}\\
 \end{center}

 \vspace{5mm}
 \centerline{{\bf{Abstract}}}
The problem of the passage of the neutral massless particle with
anomalous magnetic moment through the external electromagnetic
field is considered both in pseudoclassical and quantum
mechanics. The quantum description uses the hamiltonian in the
Foldy--Wouthuysen representation, obtained from the
pseudoclassical hamiltonian of the massive charged particle with
anomalous magnetic moment in
interaction with the external electromagnetic field using Weyl
quantization scheme.

 \vfill
 \centerline{\large Yerevan Physics Institute}
 \centerline{\large Yerevan 1995}

 \end{titlepage}
 \newpage
 \renewcommand{\thefootnote}{\arabic{footnote}}
 \setcounter{footnote}{0}

 \section{Introduction}
 \indent

     In   the   paper  of  the  authors  \cite{GG5}  in  the
pseudoclassical  approach  \cite{BM2},  when the spin degrees
of   freedom  are  described  by  Grassmann  variables,   the
canonical  quantization of the relativistic spinnig particle
with  anomalous  magnetic moment (AMM),  interacting with the
external  electromagnetic  field,   was  carried  out  in the
space--time  dimensions $D=2n$. The resulting quantum theory
is  the  theory  of  the  Dirac  particle  in  the  external
electromagnetic     field     in    the    Foldy--Wouthuysen
representation \cite{BCLFW,GG4}. The physical Hamiltonian of
the theory was found,  and in the dimension $D=4$ it is given
by the  expression
\vspace{2mm}
 \begin{eqnarray}
 \label{HAM}
 H_{\rm phys}&=&\Omega-g\kappa A_0(q, \tau)+\left[\frac{i(g-2mG)}
 {4\Omega}F_{ik}(q, \tau)\right.-\nonumber\\
 &-&\frac{ig\kappa}{2\Omega(\Omega+m)}\left(
 F_{0k}(q, \tau)\pi_j-F_{0j}(q, \tau)\pi_k\right)+\nonumber\\
 &+&\frac{i\kappa
 G}{\Omega}\left(F_{0k}(q, \tau)\pi_j-F_{0j}(q, \tau)\pi_k-
 \frac{\kappa}{\Omega+m}F_{ij}(q, \tau)\pi_i\pi_k \right.+\\
 &+&\left.\left.\frac{\kappa}
 {\Omega+m}F_{ik}(q, \tau)\pi_i\pi_j\right)\right]
 (\psi_k\psi_j-\psi_j\psi_k).\nonumber
 \end{eqnarray}
 \vspace{2mm}

      Here $g, \, m, \, -G $ are the charge,  mass and the AMM  of  the
 particle,  correspondingly,  $\kappa$ is the parameter of the
 theory,  which can  be equal to $\pm 1$ ($\kappa=+1$ corresponds
 to the presence in the theory of  the particle,  while $\kappa=-1$
 corresponds to the presence of the antiparticle),  $\tau$ is a
 parameter  along  the  trajectory of  the  particle;
 $\Omega=\sqrt{p_i^2+m^2} $,  $\pi_i=P_i-gA_i$; $A_0, \, A_i, \, F_{0k}, \,
 F_{ik} $ are components  of  the  vector-potential  and  of  the
 stress-tensor  of  the  electromagnetic  field,   correspondingly;
 $p_i, \, q_j, \,  \psi_k$ -are canonical (Newton-Wigner) variables of
 the  theory. The first two variables correspond to physical
 momentum and coordinate ,  while the spin of the particle
 is described through $\psi_k $ ($\psi_k $,  contrary
 to $p_i, \, q_j $,  are Grassmann odd variables).

     The  quantum  Hamiltonian,   corresponding  to  $H_{{\rm
phys}}$,    is   obtained   from   (\ref{HAM})   using   Weyl
prescription for operator ordering,  adapted to the presence
 of Grassmann variables in the theory \cite{BM2}. Then the operator
 $\hat{\psi}_k $  in  the  $D=4$ space-time is realized through
 Pauli matrices:
 $\hat{\psi}_k=\left(\frac{\hbar}{2}\right)^{1/2}\sigma_k$.

   Note,  that this method for the construction of the  Hamiltonian
 operator in the FW-representation  differs  from  the  usual
 method ,  when the operator in the FW -- representation  is  obtained
 from the operator in the Dirac picture using the relation
 \begin{equation}
 \label{HFW}
 \hat{H}^{FW}=\exp(-iS)\hat{H}^D\exp(iS),
 \end{equation}
 where $\hat{H}^{FW}, \, \hat{H}^D $ are Hamiltonian operators in FW
 and Dirac representations correspondingly,  $ S $  is  the
 generator of Foldy-Wouthuysen transformation \cite{FW}. As is well
 known,  for  the  free  particle the operator $ S$ can be
 construed  exactly. However in the presence of the external
field  the  generator  $S  $  (and hence $\hat{H}^{FW} $) is
known  only  in  a  series of the parameter $1/m$ \cite{FW}.
This  makes  the  transition  in  the theory to the massless
limit  impossible.  Contrary  to  this,   the  expression for
$H_{{\rm phys}} $,  as it's easy to see from (\ref{HAM}),  and
the    corresponding    quantum   Hamiltonian   operator   $
\hat{H}_{{\rm  phys}}$  are  analytic  functions of the mass
parameter $m$. Note,  that the generalization of the Foldy --
Wouthuysen  hamiltonian  to  describe  a  particle with AMM,
which  allows to set $m=0$,  was given by Sattorp and DeGroot
\cite{SG}. However their hamiltonian was correct only to the
first  order  of $e$ and contained only first derivatives of
the electromagnetic  potentials.

      Note also,  that in \cite{SPF},  where the Compton scattering
  by a proton was investigated,   the effective hamiltonian of the
 interaction of the particle with AMM with
 external electromagnetic field in the FW-representation  in  the
 $(1/m)^3$ approximation was found. Comparing that hamiltonian
 with  the  one obtained from $\hat{H}_{{\rm phys}} $  in the
 same approximation we find that  they coincide in the first order
 of $\hbar$.

      The analyticity of the expression for $H_{{\rm phys}}$,   and
hence of $\hat{H}_{{\rm phys}}$,
  with respect to the parameter $m$ allows to use (\ref{HAM})
 for $m=0$ to describe the interaction of the massless Dirac particle
 with AMM,  interacting with the external electromagnetic field and
 this problem is investigated in this paper. Namely,  we will
 consider a massless neutral particle ($m=0, \, g=0 $). This
 corresponds to the propagation of  the four component neutrino
 (with AMM equal to  $-G$)  in  the  external electromagnetic
 field. This problem attracted  attention  recently in connection
 with the attempts to  explain  the  anticorrelations between the
 number of registered solar neutrinos and the  magnetic activity
 of the sun. One of the possible explanations was based on the
 assumption that neutrino has a anomalous magnetic moment: when
 the neutrino passes through the convective outer layers of the
 sun its spin precesses in the plane,  perpendicular to the
 magnetic field of the sun. As a result some of the left neutrinos
 born in the core of the sun become right neutrinos,  that are not
 registered in the solar neutrino experiment \cite{VVO}.

   We'll return to this question after solving the problem within
 the pseudoclassical approach using the expression (\ref{HAM}) with
 $m=0, \quad g=0$.

 \section{Pseudoclassical description}
 \indent

       Now let us write down the expression for the pseudoclassical
 hamiltonian $H$,  which describes the interaction  of  the  massless
 neutral particle with AMM with the external electromagnetic field.
 From (\ref{HAM}) we find for $ g=0, \, m=0$
 \begin{eqnarray}
 \label{HNE}
 H&=&\sqrt{\vec{p}^2}+2G\left[\vec{B}\vec{S}-
 (\vec{B}\vec{n})(\vec{S}\vec{n})\right]-2\kappa G\vec{S}\cdot
 \left[\vec{E}, \vec{n}\right]=\nonumber\\
 &=&|\vec{p}|+2G\vec{S}\vec{D},
 \end{eqnarray}
 where
 \begin{equation}
 \label{VEC}
 \vec{D}=\vec{B}-(\vec{B}\vec{n})\vec{n}
-\kappa\left[\vec{E}, \vec{n}\right])
 \end{equation}

 In writing down the expressions (\ref{HNE}) the following
 notations were
 used:$E_k=F_{ok}$,  $B_i=\frac{1}{2}\varepsilon_{ijk}F_{kj}$; $S_i$
 is the pseudoclassical spin vector,  defined as
 $S_i=-\frac{i}{2}\varepsilon_{ikl}\psi_k\psi_l$,  $\vec{n}$,
 $n_i=p_i/|\vec{p}|$. Note that only transverse components of the
 fields $\vec{E}, \,  \vec{B}$ enter the interaction hamiltonian:
 $B_i^{\perp}=B_i-(\vec{B}\vec{n})n_i$,
 $\left[\vec{E}, \vec{n}\right]= \left[\vec{E}^{\perp}, \vec{n}\right]$. Thus
the $\vec{D}$ lies in the plane,  which is perpendicular to
 the momentum of the particle.

 Consider now the equations of motion of canonical variables
 $q_i, \, p_j$ and of the spin ${S_k}$
 \begin{equation}
 \label{EQ}
 \dot{q}_i=\left\{q_i, H\right\}_D=
\frac{\partial H}{\partial p_i},
\end{equation}
 \begin{equation}
 \label{EP}
 \dot{p_j}=\left\{p_j, H\right\}_D=
-\frac{\partial H}{\partial q_j},
 \end{equation}
 \begin{equation}
 \label{ES}
 \dot{S_k}=\left\{S_k, H\right\}_D.
 \end{equation}
 Here Dirac brackets are given by the relations
 $$ \left\{q_i, p_j\right\}_D=\delta_{ij}, \quad
 \left\{\psi_i, \psi_j\right\}=-i\delta_{ij}
 $$
 (all other brackets  for the variables
 $q_i, \, p_j\, \psi_k$    are equal to zero).
  Using the explicit expression for the Hamiltonian (\ref{HNE})
 we find from (\ref{EQ})-(\ref{ES})
 \begin{equation}
 \label{Q1}
 \dot{q}_i=n_i,
 \end{equation}
 \begin{equation}
 \label{P1}
 \dot{p}_j=-2GS_k\frac{\partial D_k}{\partial q_j},
 \end{equation}
 \begin{equation}
 \label{S1}
 \dot{S}_k=2 G\varepsilon_{klm} D_lS_m.
 \end{equation}
 The eq. (\ref{Q1}) reflects the fact,  that the  modulus  of  the
 velocity of the particle doesn't change and is equal to
 $|\dot{q}_i|=1 $  (in
 units $c=1$). As it follows from the equation (\ref{P1})  the
 momentum of the particle is constant in the presence of  the
 constant external electromagnetic field (note that the vector  $
 \vec{D} $
 lies in plane,  perpendicular to the momentum vector). As for the
 equation (\ref{P1}),  it describes
the precession of the pseudoclassical
 spin vector around the external "magnetic" field $\vec{D}$.

 In what follows we will consider the motion of the spinning
 particle in the constant electromagnetic field. Hence we have
 $p_i={\rm const}$. Without loss of generality the axis $q_3$ can
 be chosen along the direction of the momentum vector:
 $n_i=(0, 0, 1)$,  and the axis $q_1$ along the $\vec{D}$. In this
 coordinate system the expression (\ref{HNE}) for the
 hamiltonian takes the form
 \begin{equation}
 \label{H}
 H=|p_3|+2GS_1D_1, \quad D_1=B_1-\kappa E_2
 \end{equation}
 and the equations (\ref{S1}) are rewritten as follows:
 \begin{equation}
 \label{SE}
 \dot{S}_1=0\quad\dot{S}_2=-2GD_1S_3, \quad\dot{S}_3=2GD_1S_2
 \end{equation}
 Evidently the solution of the first equation is $S_1={\rm const}$.
 We'll write down the solution of the remaining set of equations
 in the form
 \begin{equation}
 \label{SS}
 S_2=A_2 e^{2iGD_1t}+B_2 e^{-2iGD_1t},
 \quad \quad S_3=A_3 e^{2iGD_1t}+B_3 e^{-2iGD_1t}
 \end{equation}
 where $A_2, \, B_2, \, A_3, \, B_3$ are some constants. Substituting
 (\ref{SS}) into (\ref{SE}) we find relations between this constants
 \begin{equation}
 \label{AB}
 A_3=-iA_2, \quad\quad B_3=iB_2.
 \end{equation}

 Suppose now,  that for $t=0$ the pseudoclassical spin vector
 had projections $S_2=0, \, S_3=\frac{1}{2}\hbar$.
  In quantum case this corresponds to
 the particle with positive chirality. This initial condition
 allows to find all coefficients $A_2, \, B_2, \, A_3, \, B_3$:
 \begin{equation}
 \label{ABAB}
 A_2=\frac{i}{2}, \quad B_2=-\frac{i}{2};\quad A_3=\frac{1}{2}, \quad
 B_3=\frac{1}{2}.
 \end{equation}
 Hence from (\ref{SS}) we find
 \begin{equation}
 \label{SF}
 S_2=-\frac{1}{2}\hbar \sin(2GD_1t), \quad \quad
  S_3=\frac{1}{2}\hbar \cos(2GD_1t).
 \end{equation}
 Thus the motion of the spin is a precession in plane $(q_2q_3)$
 ($S_1$- component is a constant of a motion)

 \section{Quantum description}
 \indent

 Consider now the quantum mechanical problem of the propagation
 of the neutrino in the constant electromagnetic field. The
 choice of the coordinate system is as in sect.2: the momentum of
 the particle is in direction of $q_3$,  while the axis $q_1$
 is directed along the vector $\vec{D}$. The neutrino,  having
 the momentum $\vec{p}=(0, 0, p_3)$ is moving from the region
 $q_3<0$ towards the region of the constant electromagnetic
 field,  confined  between the planes $q_3=0$ and $q_3=L$.
 The Hamilton operator of the system in the region $0<q_3<l$
 is found from (\ref{H}) and is given by
 \begin{equation}
 \label{QH}
 \hat{H}=|\hat{p}_3|+G\hbar D_1 \sigma_1,
 \end{equation}
 where the expression for the spin operator is used:
 $\hat{S}_i=(\hbar/2)\sigma_i$,  $\sigma_i$ are the Pauli
 matrices. In the regions $q_3<0$ and $q_3>L$ we have a free
 motion. Denote the wave functions in the regions $q_3<0$ ,
 $0<q_3<l$ and $q_3 >L$ by $\psi^{I}, \, \psi^{II}, \, \psi^{III}$,
 correspondingly. Since the hamiltonian doesn't depend on time
 explicitly,  the problem is reduced to finding solutions  of the
  stationary states equation
 \begin{equation}
 \label{EWF}
 \hat{H}\psi_\alpha(q_3)=E\psi_\alpha(q_3)
 \end{equation}

 Consider first the region $0<q_3<l$. The Hamilton operator
 is given by (\ref{QH}). To construct the $\psi^{II}_{\alpha}$
 we'll find first the spin eigenfunctions $\chi_{\alpha}$
  from the equation
 \begin{equation}
 \label{QSE}
 (D_1\sigma_1)_{\alpha\beta}\chi_\beta=\lambda\chi_\alpha,
 \quad \chi_\alpha=(\chi_1, \chi_2),
 \end{equation}
 where $\lambda$ the eigenvalues,  corresponding to this functions.
 Nontrivial solutions of this equation exist under the condition
 \begin{equation}
 \label{DET}
 \det|D_1\sigma_1-\lambda|=0,
 \end{equation}
  which together with the normalization condition
  $\chi_1^2+\chi_2^2=1$ leads to the following eigenvalues
  and eigenfunctions:
 \begin{eqnarray}
 \label{SPF}
 \chi_\alpha^1&=&\frac{1}{\sqrt{2}}\left(\begin{array}{c}1\\1
 \end{array}
 \right), \quad \lambda_1=D_1\nonumber\\
 \chi_\alpha^2 &=& \frac{1}{\sqrt{2}}\left(\begin{array}{c}1\\-1
 \end{array}\right), \quad \lambda_2=-D_1
 \end{eqnarray}

 We'll look for the solution of (\ref{EWF}) in the form of the
  decomposition through spin eigenfunctions
 $\chi_\alpha^i=(\chi_\alpha^1, \chi_\alpha^2)$:
 \begin{equation}
 \label{WES}
 \psi_\alpha^{II}(q_3)=\sum_{i}f_i(q_3)\chi_\alpha^i,
 \end{equation}
 where $f_i(q_3)$ are scalar functions of the coordinates.
 Substituting (\ref{WES}) into (\ref{EWF}) and taking into
 account the equation (\ref{QSE}) and the completeness
 of the set of functions $\chi_\alpha^i$,  we come to the
 equation
 \begin{equation}
 \label{QE1}
 \left(|\hat{p}_3|+G\hbar\lambda_{(i)}-E\right)f_i(q_3)=0
 \end{equation}
 The general solution of this equation is given by the
 expression
 \begin{equation}
 \label{SQE}
 f_i(q_3)=a_i e^{\frac{i}{\hbar}
p_3^{(i)}q_3}+b_i e^{-\frac{i}{\hbar}p_3^{(i)}q_3}
 \end{equation}
 $$p_3^{(i)}=E-G\hbar \lambda_{(i)}$$,
 where $a_i, \, b_i$ are some constants.
 Substituting now (\ref{SQE}) in (\ref{WES}) we come to the following
 expression for the wave function $\psi_\alpha^{II}(q_3)$:
 \begin{equation}
 \label{WFII}
 \psi_\alpha^{II}(q_3)=\sum_{i=1, 2}\left(
 a_i e^{\frac{i}{\hbar}p_3^{(i)}q_3}+
b_i e^{-\frac{i}{\hbar}p_3^{(i)}q_3}\right)\chi_\alpha^i
 \end{equation}

 In the region $q_3<0$ both incident and reflected waves are present.
 The corresponding wave function $\psi_\alpha^{I}(q_3)$,
 presented as a decomposition through spin eigenfunctions
 $ \chi_\alpha^i$ has the form
 \begin{eqnarray}
 \label{WFI}
 \psi_\alpha^{I}(q_3) & = &\frac{1}
{\sqrt{2}}\chi_\alpha^1 e^{\frac{i}{\hbar}p_3 q_3}+
 \frac{1}{\sqrt{2}}\chi_\alpha^2
e^{\frac{i}{\hbar}p_3 q_3}+
 \chi_\alpha^1 A_1e^{-\frac{i}
{\hbar}p_3 q_3}+\chi_\alpha^2 A_2e^{-ip_3 q_3}=\nonumber\\
 & = & e^{\frac{i}{\hbar}p_3 q_3}
\left(\begin{array}{c}1\\0\end{array}\right)+
 e^{-\frac{i}{\hbar}p_3 q_3}
\frac{1}{\sqrt{2}}\left(\begin{array}{c}A_1+A_2\\
 A_1-A_2\end{array}\right),
 \end{eqnarray}
 where $A_1, \, A_2$ are some constants. The first summand in the
 second line of (\ref{WFI}) corresponds to the wave of definite
 polarization (as it was mentioned earlier,  the neutrinos born
  in the core of the sun are left handed)

  In the region $q_3>L$ only the outgoing wave ,  which moves
 in the direction of the incident wave,  is present:
 \begin{eqnarray}
 \label{WFIII}
 \psi_\alpha^{III}(q_3)&=&
 \chi_\alpha^1 B_1e^{\frac{i}{\hbar}p_3 q_3}+
 \chi_\alpha^2 B_2e^{\frac{i}{\hbar}p_3 q_3}=\nonumber\\
 &=&e^{\frac{i}{\hbar}p_3 q_3}\frac{1}{\sqrt{2}}
 \left(\begin{array}{c}B_1+B_2\\
 B_1-B_2\end{array}\right)=
 e^{\frac{i}{\hbar}p_3 q_3}\chi_\alpha^{III}
 \end{eqnarray}
 where $B_1, \, B_2$ are constants.

 The constants $a, \, b, \, A, \, B $ are defined from the condition of
 continuity of $\psi$ and $d\psi/dt$ at the planes $q_3=0$ and
 $q_3=L$. Because of the orthogonality of the functions
 $\chi_\alpha^1, \, \chi_\alpha^2$,  the corresponding set of
 equations is divided into two
subsets of equations,  each containing
 only one of the functions  $\chi_\alpha^1, \, \chi_\alpha^2$.
 Solving these equations we come to the following values for the
 constants:
 \begin{equation}
 \label{CA1}
 A_1=-\frac{i\sqrt{2}G\hbar D_1\sin(\frac{1}{\hbar}p_3^{(1)}L)}
 {(E+p_3^{(1)})\left[
 e^{-\frac{i}{\hbar}p_3^{(1)}L}-
\frac{(G\hbar D_1)^2}{(E+p_3^{(1)})^2}
 e^{\frac{i}{\hbar}p_3^{(1)}L}\right]}
 \end{equation}
 \begin{equation}
 \label{CA2}
 A_2=-\frac{i\sqrt{2}G\hbar D_1\sin(\frac{1}{\hbar}p_3^{(1)}L)}
 {(E+p_3^{(2)})\left[
 e^{-\frac{i}{\hbar}p_3^{(2)}L}-
\frac{(G\hbar D_1)^2}{(E+p_3^{(2)})^2}
 e^{\frac{i}{\hbar}p_3^{(2)}L}\right]}
 \end{equation}
 \begin{equation}
 \label{CB1}
 B_1=\frac{\sqrt{2}}{2}\frac{e^{-iG D_1 L}}{1-\frac{i(G \hbar D_1)^2}
 {2Ep_3^{(1)}}\sin(\frac{1}{\hbar}p_3^{(1)}L)
 e^{\frac{i}{\hbar}p_3^{(1)}L}}
 \end{equation}
 \begin{equation}
 \label{CB2}
 B_2=\frac{\sqrt{2}}{2}\frac{e^{iG D_1 L}}{1-\frac{i(G \hbar D_1)^2}
 {2E p_3^{(2)}}\sin(\frac{1}{\hbar}p_3^{(2)}L)
 e^{\frac{i}{\hbar} p_3^{(2)}L}}
 \end{equation}

 As it was shown in \cite{VVO},  the value of
the quantity $G\hbar D_1/E$
 is small for the real electromagnetic fields and medium energies
 of the neutrino. Hence the amplitudes of the reflected waves,
 as one can see from (\ref{CA1}) and (\ref{CA2}),  are also small.
 As for the outgoing wave,  taking into account the smallness of the
 named parameter,  we find from (\ref{CB1}) and (\ref{CB2}) that
 \begin{equation}
 \label{PW}
 B_1=\frac{\sqrt{2}}{2}e^{-iG D_1 L}, \quad
 B_2=\frac{\sqrt{2}}{2}e^{iG D_1 L},
 \end{equation}
 which in their turn bring as to the following expression for
 $\psi_\alpha^{III}$ (see (\ref{WFIII}):
 \begin{equation}
 \label{PWF}
 \psi_\alpha^{III}=e^{\frac{i}{\hbar}p_3 q_3}
 \left(\begin{array}{c}\cos(G D_1 L)\\
 -i\sin(G D_1 L)
 \end{array}
 \right)
 \end{equation}

 To compare this result with the pseudoclassical one we calculate
 the mean projection of the spin of the outgoing wave on the $q_2, \, q_3 $
axes
 \begin{equation}
 \label{SM}
 <S_2>=\chi_\alpha^{+\, III} \frac{\hbar}{2}\sigma_2 \chi_\alpha^{III}
 = - \frac{\hbar}{2}\sin(2G D_1 L),  \quad
 <S_3>=\chi_\alpha^{+\, III} \frac{\hbar}{2}\sigma_3 \chi_\alpha^{III}
 = \frac{\hbar}{2}\cos(2G D_1 L)
 \end{equation}

 If we now insert in (\ref{SF}) $t=L$,  we'll find a coincidence
of the results of the quantum and pseudoclassical description
of the rotation of the spin of the massless neutral particle
in the external electromagnetic field.

 \end{document}